\documentclass[aps,prl,twocolumn,showpacs,superscriptaddress,floatfix]{revtex4-1}
\usepackage{graphicx,amsmath,amssymb}
\usepackage[usenames]{color}
\usepackage[dvipsnames]{xcolor}
\usepackage[unicode=true,pdfusetitle,
 bookmarks=false,
 breaklinks=false,pdfborder={0 0 1},backref=false,colorlinks,urlcolor=Black, linkcolor=Red,citecolor=BrickRed]
 {hyperref}
\def\I{{\mathcal I}}
\def\l{\lambda}

\def\rd{{\textrm{d}}}
\def\om{\omega}
\def\D{\Delta}

\begin{document}

\title{Critical Quantum Metrology in the Non-Linear Quantum Rabi Model
}

\author{Zu-Jian Ying }
\email{yingzj@lzu.edu.cn}
\affiliation{School of Physical Science and Technology,
Lanzhou University, Lanzhou 730000, China}

\author{Simone Felicetti }
\email{felicetti.simone@gmail.com}
\affiliation{Institute for Complex Systems, National Research Council (ISC-CNR), 00185 Rome, Italy
}

\author{Gang Liu }
\affiliation{School of Physical Science and Technology,
Lanzhou University, Lanzhou 730000, China}

\author{Daniel Braak }
\email{daniel.braak@physik.uni-augsburg.de}
\affiliation{EP VI and Center for Electronic Correlations and Magnetism, University of Augsburg, 86135 Augsburg, Germany}

\begin{abstract}
  The quantum Rabi model (QRM) with linear coupling between light mode and qubit exhibits the analog of a second order phase transition for vanishing mode frequency which allows for criticality-enhanced quantum metrology in a few-body system. We show that the QRM including a non-linear coupling term exhibits much higher measurement precisions due to its first order like phase transition at \emph{finite} frequency, avoiding the detrimental slowing-down effect close to the critical point of the linear QRM. When a bias term is added to the Hamiltonian, the system can be used as a fluxmeter or magnetometer if implemented in circuit QED platforms.
\end{abstract}
\pacs{ }
\maketitle


\section{Introduction }

The high susceptibility developed by critical systems~\cite{Huang,Tauber} in proximity of phase transitions is a compelling resource for metrology and sensing. For example, relevant  scientific and technological applications of critical systems are bubble chambers~\cite{bolometer_review} and transition-edge sensors~\cite{irwin_transition-edge_2005}. However, even when these devices have a quantum working principle, they follow a classical sensing strategy. Yet, it is well known that quantum properties such as squeezing and entanglement can be used to outperform any classical sensing protocol~\cite{Degen:17}.  As systems in proximity of quantum phase transitions~\cite{Sachdev} are expected to have a highly nonclassical behavior, it is natural to analyze critical  systems with a quantum-metrology perspective. In the last decade, various theoretical works have introduced different protocols able to leverage quantum critical phase transitions to achieve a fundamental advantage over classical sensing strategies~\cite{zanardi_quantum_2008, invernizzi2008Optimal,ivanov_adiabatic_2013, bina_dicke_2016, fernandez-lorenzo_quantum_2017,tsang_quantum_2013,macieszczak_dynamical_2016}. However, an often-neglected fundamental hindrance limits the performances of critical quantum sensors: The diverging susceptibility is counterbalanced by the critical slowing down, which implies an extremely long protocol duration time. Only very recently it has been shown that, counterintuitively, even in presence of the critical slowing down the optimal limit of precision can be achieved~\cite{rams_at_2018}. Indeed, under standard assumptions critical protocols can achieve the Heisenberg scaling ---  a quadratic growth of parameter-estimation precision --- both with respect to the number of probes and with respect to measurement time. Furthermore, a recent theoretical work~\cite{garbe2020} demonstrated that the optimal limits of precision can be achieved using finite-component phase transitions~\cite{bakemeier2012quantum,Ashhab2013,Hwang2015PRL,Ying2015,LiuM2017PRL,Puebla:16,Puebla:17,Hwang:18,Zhu:20,Puebla:20b,
Ying-2021-AQT,Ying-gapped-top,Liu2021AQT}, which are criticalities that take place in quantum optical systems where the thermodynamic limit is replaced by a scaling of the system parameters~\cite{LiuM2017PRL,Casteels2017,Bartolo2016Exact,Minganti2018,peng_unified2019,felicetti2020universal,kewming2022diverging}. Critical quantum sensors can then be implemented also with controllable small-scale quantum devices, without requiring the control of complex many-body systems. These results have prompted an intense research effort dedicated to designing efficient protocols~\cite{Ivanov2020,Chu2021,gietka2021,hu2021,liu2021,ilias2021criticality, frerot_quantum_2018, heugel2020_quantum, Mirkhalaf2020, Wald2020, Ivanov_2020steady, Salado2021, Niezgoda2021, mishra2021integrable,garbe2021exponential,Gietka22_squeeze,Gietka2022understanding} in terms of high estimation precision and limited measurement time, and which can be implemented with experimentally-feasible operations. Practical applications in quantum magnetometry and superconducting-qubit readout have also been proposed~\cite{di2021critical}.

Critical quantum metrology protocols can be divided in two main approaches. The \emph{static} approach~\cite{zanardi_quantum_2008, invernizzi2008Optimal,ivanov_adiabatic_2013, bina_dicke_2016, fernandez-lorenzo_quantum_2017, rams_at_2018, heugel2020_quantum, Mirkhalaf2020, Wald2020, Ivanov_2020steady, Salado2021, Niezgoda2021, mishra2021integrable,garbe2020,Montenegro2021PRL} consists in bringing the system in an equilibrium state that depends  on an external perturbation (such as a magnetic field). Such equilibrium states can be represented by the ground state reached during an adiabatic sweep, or by the steady-state achieved after a long-time evolution in a driven-dissipative setting.
When the system is brought in proximity of the phase transition, one can get a very precise estimate of the parameter by measuring an observable on the equilibrium state. Differently, the \emph{dynamical} approach~\cite{tsang_quantum_2013,macieszczak_dynamical_2016,Chu2021} consists in preparing the probe in a known state to then apply the perturbation and monitor the system time evolution, which can also have a critical dependence on the system parameters.
 Recent results obtained with spin systems and finite-component transitions suggest~\cite{rams_at_2018} that the dynamical and equilibrium approaches have a similar scaling of the estimation precision in the thermodynamic (or parameter-scaling) limit. However, the dynamical approach can achieve a constant factor advantage over static protocols~\cite{Chu2021}, and it can allow super-Heisenberg scaling in collective light-matter interaction models~\cite{Gietka2022understanding}. For fully-connected models, it has recently been shown that a continuous connection~\cite{Garbe_2022_connect} can be drawn between the static and dynamical approaches, identifying universal time-scaling regimes.

In the design of critical quantum sensing protocols a variety of physical models have been considered, such as many-body spin systems~\cite{rams_at_2018}, ensemble of emitters coupled to cavity modes~\cite{bina_dicke_2016}, single atom-cavity models~\cite{garbe2020,Chu2021}, and nonlinear quantum resonators~\cite{di2021critical}. So far, except for few exceptions, most studies have focused on the parameter regime defined by thermodynamic or parameter-scaling limits, where an effective analytical description can be derived. When considering finite-component phase transition, the most widely studied case is the quantum Rabi model (QRM)~\cite{Rabi:37,Jaynes:63,Braak:16}, composed of a  two-level atom coupled to a single quantum harmonic mode. This model undergoes a second-order critical phase transition in the slow-resonator limit~\cite{bakemeier2012quantum,Ashhab2013,Hwang2015PRL,Ying2015,LiuM2017PRL,Ying-2021-AQT,Ying-gapped-top,Liu2021AQT}, where the frequency of the mode and the coupling strength are sent to zero with a given scaling law. Focusing on the scaling limit one can obtain interesting results on the  growth of the estimation precision in terms of fundamental resources such as the size of probe systems or photon number but, to assess the actual precision of practical protocols, finite values of the parameters must be considered.

In this work we propose quantum critical sensing protocols based on a generalization of the quantum Rabi model which includes a non-linear (two-photon) coupling term and a transversal spin bias. The linear and non-linear interactions lead to a ground state whose dependence on the linear coupling is much stronger at the critical value, entailing the equivalent of a first order quantum phase transition.
We consider the static approach where an adiabatic sweep is used to bring the system in proximity of criticality and we perform a numerical analysis which is not limited to the
scaling regime. We show that adding the nonlinear coupling and the bias improves the protocol efficiency in different ways: 1) Higher estimation precision, as measured by an increase of the quantum Fisher information; 2) Faster adiabatic sweep and so shorter protocol duration time, due to the larger energy gap for finite values of physical parameters; 3) An extended range of the efficient sensing region, as the position of the critical point can be tuned in the space of parameters; 4) a less challenging requirement on the implementation of the slow-resonator limit. The considered model can be feasibly implemented with atomic~\cite{Bertet-Nonlinear-Experim-Model-2005, Felicetti2015-TwoPhotonProcess,Puebla17_p,Cong20_sel} and solid-state~\cite{Two_ph_circuit,2ph_USC,Munnoz20} quantum devices with nowadays technology.

\begin{figure}[t]
\includegraphics[width=1.0\columnwidth]{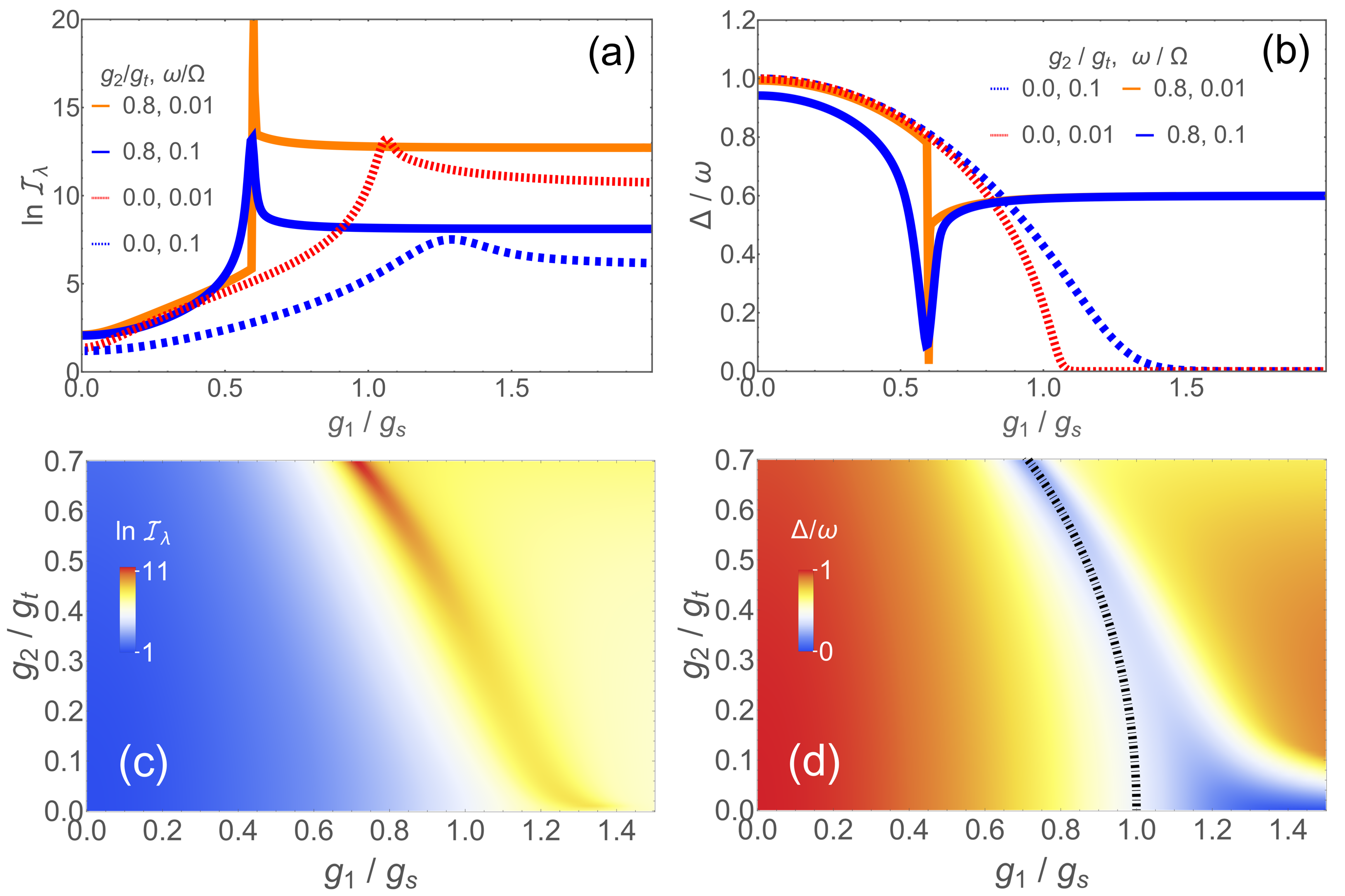}
\caption{(color online) (a) $\ln \I_\l$ for $\l=g_1/\Omega$ and different $g_{2}$: $g_{2}=0$ with $\omega /\Omega =0.01$ (dotted red line) and $\omega /\Omega =0.01$ (dashed blue line). $g_{2}/g_{\mathrm{t}}=0.8$
  with $\omega /\Omega =0.01$ (solid orange line) and $\omega/\Omega =0.01$ (solid blue line).
  (b) Gap $\Delta /\omega$ for the same parameters as in (a).
  (c) $\ln \I_\l$ in the $g_{1}$/$g_{2}$ plane for  $\omega /\Omega =0.1$.
  (d) Gap $\Delta /\omega$ for the same parameters as in (c). The dashed-dotted line
  represents the phase boundary given in \eqref{g1cSemiclassical}.
}
\label{fig-Compare-orders}
\end{figure}

\section{Model }

The non-linear QRM with bias
is described by the
Hamiltonian\cite{Ying2020-nonlinear-bias,Bertet-Nonlinear-Experim-Model-2005}
\begin{eqnarray}
  H&=&H_{0}+H_{t}+H_{\epsilon }, \label{ham}\\
H_{0} &=&\omega a^{\dag }a+\frac{\Omega }{2}\sigma _{x}+g_{1}\sigma _{z}{
  (a^{\dag }+a)},
\nonumber\\
H_{t} &=&g_{2}\sigma _{z}{\left[ (a^{\dag })^{2}+a^{2}
\right] ,\quad }H_{\epsilon }=-\epsilon \sigma _{z},  \nonumber
\end{eqnarray}
where $\sigma _{x,y,z}$ are Pauli matrices and $a^{\dagger }(a)$ creates
(annihilates) a bosonic mode with frequency $\omega$. The term proportional to $\Omega $ corresponds to  tunneling between two states of the flux qubit in circuit QED implementations~\cite{flux-qubit-Mooij-1999,Two_ph_circuit}, or to electronic-level splitting in trapped-ion implementations~\cite{Felicetti2015-TwoPhotonProcess}. The strengths of
linear and nonlinear couplings are denoted by $g_{1}$ and $g_{2}$
respectively.
The bias
term $H_{\epsilon }$ can be easily tuned by a bias current or by a static magnetic or electric field, depending on the implementation. In the slow-resonator limit $\omega\rightarrow 0$, the model exhibits the analogue of both second-order and first-order phase transitions  as the thermodynamic limit in a many-body system is simulated here through the infinitesimal level spacing \cite{Ying2020-nonlinear-bias}. At finite frequencies the discontinuities in the parameter dependence of expectation values are rounded off but show remnants of criticality.
It should be noted that the parity symmetry of the linear QRM ($H_{0}$)
is broken by $H_{t}$ and $H_{\epsilon }$.

\begin{figure}[t]
\includegraphics[width=0.95\columnwidth]{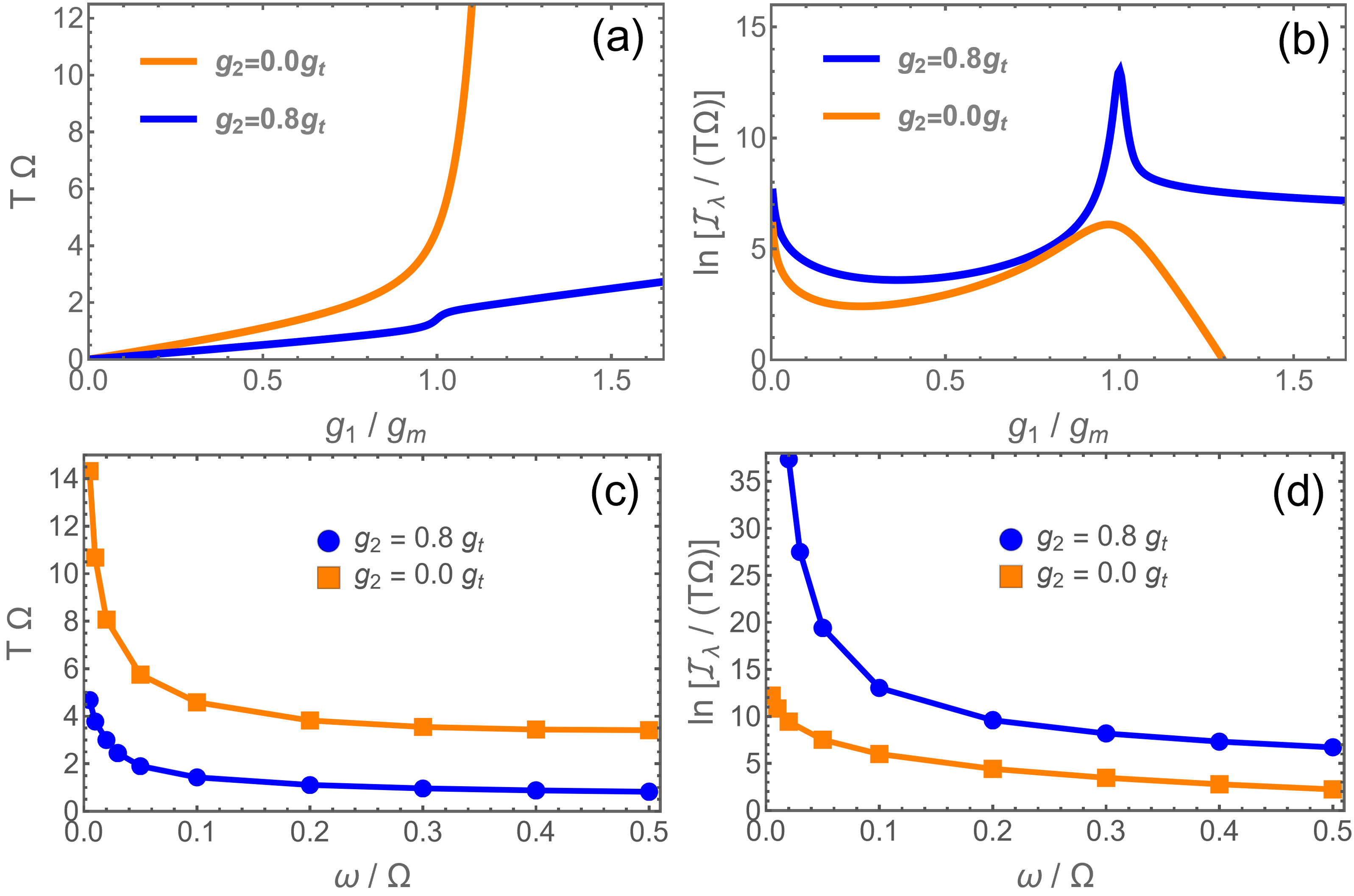}
\caption{(color online) (a) Time $T$  needed to prepare the ground state of the system at $g_1$ for different $g_2$:
$g_2=0$ (orange) and $g_2=0.8g_{\mathrm{t}}$ (blue), for $\omega =0.1 \Omega$. $g_m $ denotes the coupling $g_1$ with
  maximal Fisher information $\I_\l$.
  (b) $\ln(\I_\l/(T\Omega))$ for $g_2=0$ (orange) and $g_2=0.8g_{\mathrm{t}}$ (blue), for the same parameters as in (a).
  (c) Dependence of $T$ at $g_m$ on the mode frequency $\omega$ for the two values for $g_2$ shown in (a), (b).
  (d) $\ln(\I_\l/(T\Omega))$ at $g_m$  as function of $\omega$ for the same parameters as in (c).}
\label{fig-Compare-orders2}
\end{figure}

\section{Relation between transition order and accuracy}

To discuss the difference between the linear ($g_2=0$) and non-linear ($g_2\neq 0$) cases for quantum metrology, let us
start with zero bias $\epsilon =0$.
 The model has a phase transition in the low
 frequency limit $\omega \rightarrow 0$ at the critical point $g_1=\left\vert g_{1c}\right\vert =\left\vert g_{\mathrm{s}}\right\vert \sqrt{1-g_{2}^{2}/g_{\mathrm{t}}^{2}}$ \cite{Ying2020-nonlinear-bias,Ying-2018-arxiv} with $g_{\mathrm{s}}=\sqrt{\omega \Omega }/2$ \cite{Ashhab2013,Ying2015},
 and
 $g_{\mathrm{t}}=\omega /2$
 is the critical value of $g_2$ beyond which the Hamiltonian \eqref{ham}
 is no longer self-adjoint and becomes unphysical \cite{Felicetti2015-TwoPhotonProcess,e-collpase-Lo-1998,e-collpase-Duan-2016,e-collpase-Garbe-2017,CongLei2019,Braak2022}.
 The transition in this limit is second-order-like at $g_{2}=0$ \cite{Ashhab2013,Ying2015,Hwang2015PRL,Ying-2021-AQT,LiuM2017PRL,Ying-gapped-top,Liu2021AQT} and first-order-like at  finite $g_{2}$ \cite{Ying2020-nonlinear-bias,Ying-2018-arxiv}.
 The precision (signal-to-noise ratio) of any experimental
 estimation of one of the parameters $\l$ in \eqref{ham}  is bounded by
 $\I_\l^{1/2}$ \cite{Cramer-Rao-bound}, where $\I_\l$ is the quantum Fisher information \cite{Cramer-Rao-bound,Taddei2013FisherInfo,RamsPRX2018}, which takes the following form for pure states
\begin{equation}
\I_\l(|\psi\rangle) =4\left( \langle \psi ^{\prime }(\l) |\psi
^{\prime }(\l) \rangle -\left\vert \langle \psi ^{\prime }(\l) |\psi(\l) \rangle \right\vert ^{2}\right),
\label{qfi}
\end{equation}
where $^\prime$ denotes the derivative of the ground state (GS) $|\psi(\l)\rangle$ of
$H$ in \eqref{ham} with respect to $\l$.
Obviously, a higher QFI means higher measurement precision.

The Hamiltonian $H$ has several parameters that can drive a phase transition. Let
us begin with the linear coupling $g_1$ and set $\l=g_1/\Omega $
with $\Omega $ fixed.  $\l$ and $\I_\l$ are thus
dimensionless. In Fig.~\ref{fig-Compare-orders}(a) we compare the QFI for first and second order scenarios, as calculated
with exact diagonalization \cite{Ying2020-nonlinear-bias}, by plotting $\ln \I_\l$. The dashed lines illustrate the second order case  $g_{2}=0$ for two different $\omega$. One sees that the variation of the ground state with $\l$ and therefore the maximal value of $\I_\l$ becomes larger for
smaller frequencies.
The QFI for comparatively large $\omega =0.1\Omega $ shows a broad peak
shifted away from the critical point $g_{\mathrm{s}}$ for $\omega=0$ due to
the finite GS extension at  finite frequency \cite{Ying2015}. The peak
becomes sharper at lower frequency and tends to  diverge in the limit $\omega\rightarrow 0$, as indicated by the dotted red line with $\omega=0.01\Omega $.
At  finite frequency the QFI does
not diverge for $g_2=0$. The situation changes profoundly for non-zero $g_2$.
The GS wave function behaves much more singular even for $\omega =0.1\Omega$
(blue solid line), leading to a narrow peak in $\ln \I_\l$. Naturally, the maximal QFI is even higher for smaller frequency. By comparing the solid blue and dashed orange lines, we see that the same measurement precision can be obtained if $g_2\neq 0$ as in the model with $g_2=0$, although the mode frequency is an order of magnitude larger. Obviously, the presence of the non-linear term in $H$ simplifies the requirement to implement the slow-resonator limit.

These features of the QFI can be understood by comparing the behavior of the gap $\Delta$ between GS and first excited state when tuning through the phase transition, shown in Fig.~\ref{fig-Compare-orders}(b). For $g_2=0$, the scaled gap $\Delta/\omega$
goes to zero for $g_1\gtrsim g_s$.
The transition becomes continuous with $\D/\omega=0$ for $g_1\ge g_s$ in the limit
$\omega\rightarrow 0$, typical for a second order transition. Likewise, the GS wave function changes smoothly close to $g_s$ leading to the lower values for the QFI.
The closing of the gap means that the dynamical time scale $\Delta^{-1}$ diverges in approaching the critical coupling which means that an adiabatic sweep through $g_s$ would be extremely slow for $\omega\approx 0$. This problem will be addressed in the next section.

On the other hand, the gap stays always finite for $g_2\neq 0$ due to the broken parity symmetry \cite{Ying2020-nonlinear-bias},
although it changes very fast close to the critical point, even for large $\omega$, and resembles therefore a first order transition. This explains the higher QFI in the non-linear case.
The QFI and the gap as function of  $g_{1}$ and $g_{2}$  are shown in the colorplots of Fig.~\ref{fig-Compare-orders}(c,d) for $\omega =0.1\Omega $.
A larger $g_{2} $ means  higher maximal QFI, which increases dramatically if $g_2$ reaches $\sim 0.6 g_s$. In Fig.~\ref{fig-Compare-orders}(c) we plot only up to $g_2=0.7g_{\rm t}$ as the maximal QFI for larger $g_2\approx g_{\rm t}$ would be out of scale. For these values the system is close to the point of spectral collapse \cite{Felicetti2015-TwoPhotonProcess,e-collpase-Garbe-2017,Braak2022,CongLei2019} where part of the discrete spectrum becomes continuous. Although this regime may not be easily realizable, we see that it has by far the greatest potential with regard to quantum metrology.

\begin{figure}[t]
\includegraphics[width=1.0\columnwidth]{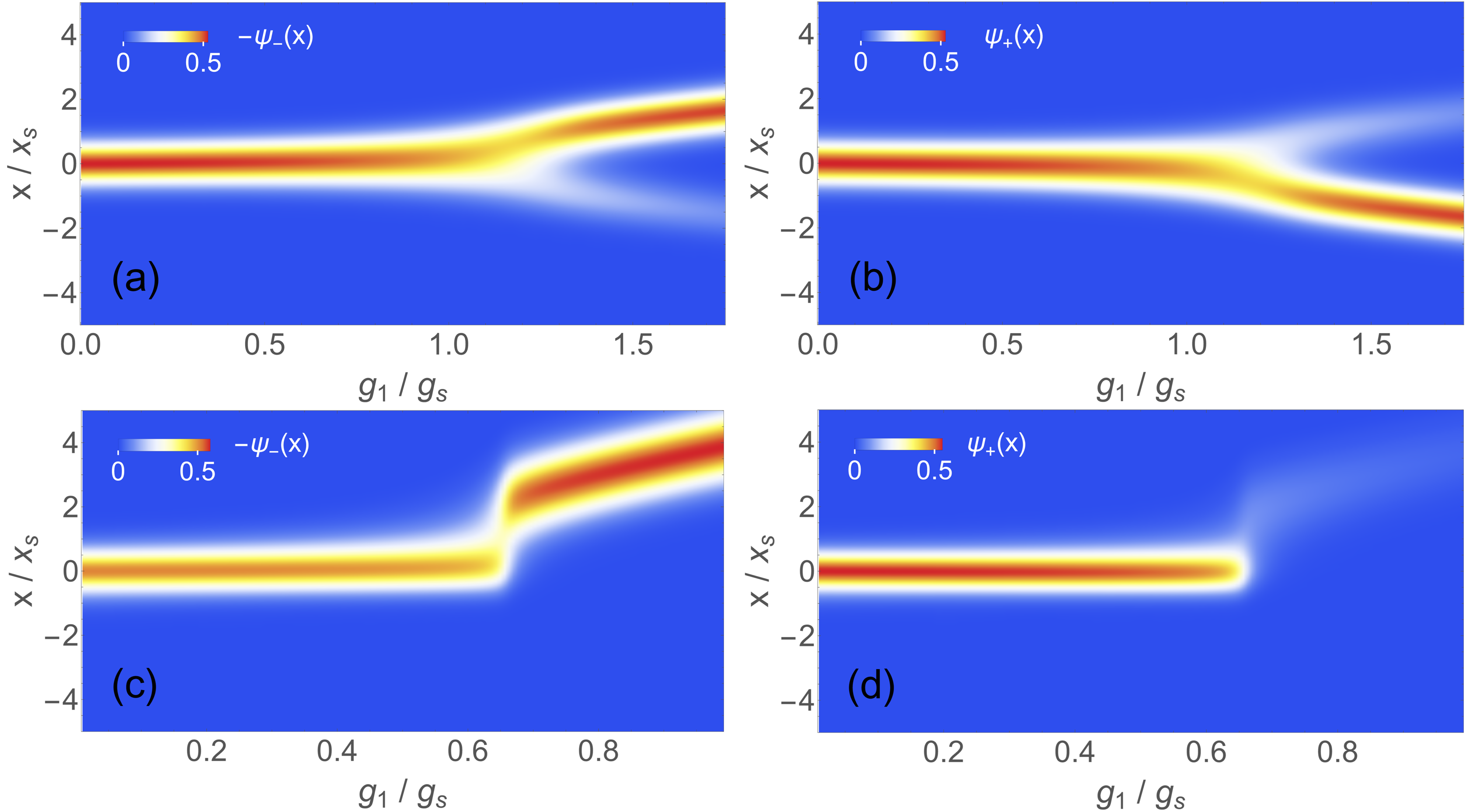}
\caption{(color online) Ground state wave function $\psi_\pm(x)$ for spin component $+$ or $-$ as function of $g_{1}$
  at $\omega /\Omega =0.1$:
  (a) $-\psi _{-}(x)$ for $g_2=0$,
  (b) $\psi _{+}(x)$ for $g_2=0$,
  (c) $\psi_-(x)$ for $g_{2}=0.75g_{\mathrm{t}}$ and
  (d) $\psi _+(x)$  for $g_{2}=0.75g_{\mathrm{t}}$.
Here $x_s=\sqrt{2}g_{\rm s} / \omega$.
}
\label{fig-WaveF}
\end{figure}

\section{Preparation Time}

To estimate the time needed to prepare the ground state of the system by adiabatic sweep from $\l=0$ to the intended sensing value $\l_{s}$, we may use the condition  $\rd\l/\rd t \ll \D(\l)$ where $\D(\l)$ is the energy gap between ground state and first excited state (see supplemental material in~\cite{garbe2020}). In this way we obtain a lower bound for the
preparation time
\begin{equation}
T(\l_s)  \ge \int_{0}^{\l_s}\frac{1}{\Delta(\l)}\rd\l.
\end{equation}
In our present case we have $\l=g_1/\Omega$.
In Fig.~\ref{fig-Compare-orders2} we compare the preparation times for  pure linear coupling $g_2=0$ (second order transition) and non-linear coupling (first order transition) at the experimentally feasible frequency ratio $\om/\Omega=0.1$. While the preparation time seems to diverge at the critical point (which is also the point of maximal QFI) due to critical slowing down in the first case, it stays low in the second. In panel (b) of Fig.~\ref{fig-Compare-orders2} we plot the logarithm of $\I_\l/(T\Omega)$, a figure of merit to assess the practicability of the sensing protocol. Around the coupling with maximal accuracy, $g_m$, the system with non-linear coupling exhibits exponential advantage compared to the linear one. In panels (c) and (d) $T$ and $\ln(\I_\l/(T\Omega))$ taken at $g_m$ are shown as function of $\om/\Omega$. The preparation time rises if one approaches the low frequency limit for linear and non-linear coupling alike because the phase transition features become more pronounced and the gap in the critical region shrinks. Still, one can see that the preparation time in the presence of non-linear coupling is much lower than without it.  For values above $\om/\Omega\sim 0.2$ the time does not change much in both cases. Likewise, the ``effective accuracy'' as measured by $\ln(\I_\l/(T\Omega))$ drops slowly for larger values of $\om$, while the non-linear system keeps an exponentially higher precision.

\section{Behavior of the wave function}

 As mentioned above, the high sensitivity of quantum metrology results
from the sudden change of the GS wave function $|\psi\rangle$ in the vicinity of the critical point. In
Fig. \ref{fig-WaveF}(a,b) we show
the components of $|\psi\rangle=(\psi_+(x),\psi_-(x))^T$ in position space
for $g_{2}=0$ and as function of $g_1$. The frequency $\om$ is relatively large ($\om/\Omega=0.1$) so that the transition is smeared out.
Below $g\approx g_s$ both spin components of $|\psi\rangle$ are centered
around $x=0$ which corresponds unbroken left/right-symmetry. Around $g_s$, the upper component is displaced to the left and the lower component to the right. This does not mean that the parity symmetry of the model with Hamiltonian $H_0$ is broken for $g > g_s$, because the parity operator $e^{i\pi a^\dagger a}\sigma_x$ acts in both spin and position space.
Nevertheless, the change of the GS wave function in position space is the analogue of a symmetry breaking quantum phase transition in the QRM. For vanishing non-linear coupling $g_2$, the change in both components is
smooth as seen in panels (a) and (b) of Fig.~\ref{fig-WaveF}.
The situation is quite different if the non-linear coupling is turned on: For non-zero $g_2$ which breaks the parity symmetry of $H_0$, we have essentially the same behavior of $\psi_\pm(x)$ for $g<0.66g_s$ as in the linear case. But at $g\approx 0.66 g_s$, the wave functions change abruptly: Basically the whole weight is transferred to the right and lower branch $\psi_-(x)$ and the parity symmetry is strongly broken. Of course, this is no symmetry breaking in the usual sense because parity is broken already on the Hamiltonian level. The fast change of $|\psi(x)\rangle$ in tuning through the transition region is responsible for the large QFI while the gap to the first excited state remains always non-zero.

\begin{figure}[t]
\includegraphics[width=1.0%
\columnwidth]{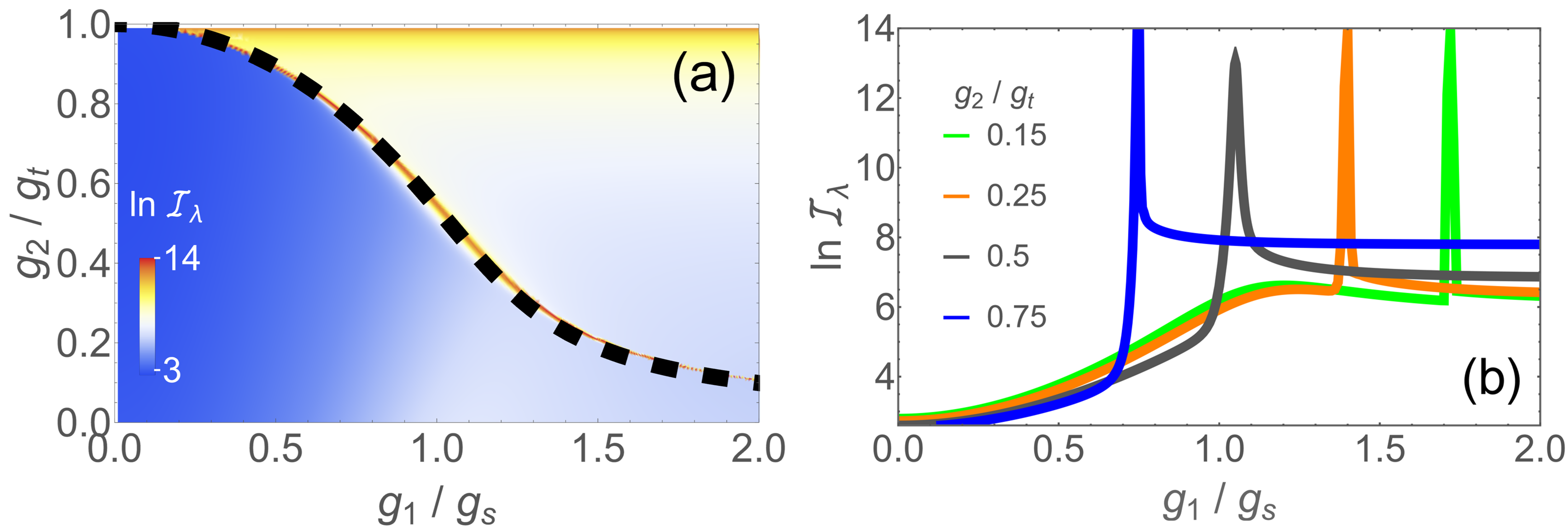}
\caption{(color online) (a) $\ln I_\l$ in the $g_1$/$g_2$ plane at fixed
  frequency $\omega/\Omega =0.1$ and finite bias  $\epsilon =0.1\Omega$.
  The dashed line denotes the analytic phase boundary  where $\I_\l$ is maximal. (b) $\ln \I_\l$ as function of $g_1$ for
different non-linear couplings: $g_2/g_{\mathrm{t}}=0.75$ (blue), $0.5$
(dark gray), $0.25$ (orange), $0.1 5$ (green). }
\label{fig-Global-sensing}
\end{figure}

\section{extended range quantum sensing}

Up to now, we have set the bias $\epsilon$ to zero to demonstrate the main differences between linear and non-linear models with regard to quantum sensing. From Fig.~\ref{fig-Compare-orders}~(c), (d) we see that in varying $g_2$ between zero and $0.7g_{\rm t}$, we can drive the critical coupling $g_1$ which is the quantity to be measured, from  $g_s$ to lower values $\sim 0.7g_s$, thus extending the range of couplings which can be measured with an accuracy enhanced by criticality. A much larger region of couplings becomes available if the bias $\epsilon$ is varied as well which can be easily achieved e.g. in circuit QED platforms.

In Fig.~\ref{fig-Global-sensing}~(a) we show the QFI in the $g_{1}$/$g_{2}$ plane in the
presence of  finite bias $\epsilon =0.1\Omega $ at $\omega =0.1\Omega $.
The phase transition occurs along the
thin red line indicating the sharp maximum of the QFI. The transition line is accurately given by a semi-classical calculation (black dashed line) in closed form as \cite{Ying2020-nonlinear-bias}
\begin{align}
  g_{1c}^\epsilon
  &=g_{\mathrm{s}}\left[ 1+\frac{g_{\mathrm{t}}\epsilon}{g_{2}\Omega}\right]
\sqrt{1-g_{2}^{2}/g_{\mathrm{t}}^{2}},
\label{g1cSemiclassical} \\
\epsilon _{c} &=\frac{g_{2}}{g_{\mathrm{t}}}\left[\frac{
g_{1}}{g_{\mathrm{s}}\sqrt{1-g_{2}^{2}/g_{\mathrm{t}}^{2}}}-1\right]\Omega.  \label{hzcSemiclassical}
\end{align}
This phase boundary cuts no longer the $x$-axis at a finite value of $g_1$ as in  Fig.~\ref{fig-Compare-orders}~(c), but allows for arbitrary large critical values of $g_1$ for non-zero $g_2$. The range of accessible couplings is extended therefore also to values above $g_s$. In this way, the whole range $0<g_1<\infty$ can be measured with enhanced precision if the couplings $g_2$ and $\epsilon$ are properly tuned.

The QFI has a peak along the phase boundary. This is shown in Fig.~\ref{fig-Global-sensing}~(b) for various values of $g_{2}$.
One may notice a shallow maximum of the lines for $g_{2}/g_{\mathrm{t}}=0.15$ and $0.25$, around $g_{1}/g_{\mathrm{s}}=1.2$ before the sharp peak associated with the first order transition.
It originates in a second order transition because the system is located in the vicinity of a tricritical point \cite{Ying2020-nonlinear-bias}. However, these maxima lead only to marginal enhancement of QFI and play no role in the optimal measurement protocol.
The contrast between the shallow maximum and the sharp peak for the same $g_2$ demonstrates again the much higher measurement accuracy made possible by a first-order-like transition compared to a transition of second order.

\begin{figure}[t]
\includegraphics[width=1.0%
\columnwidth]{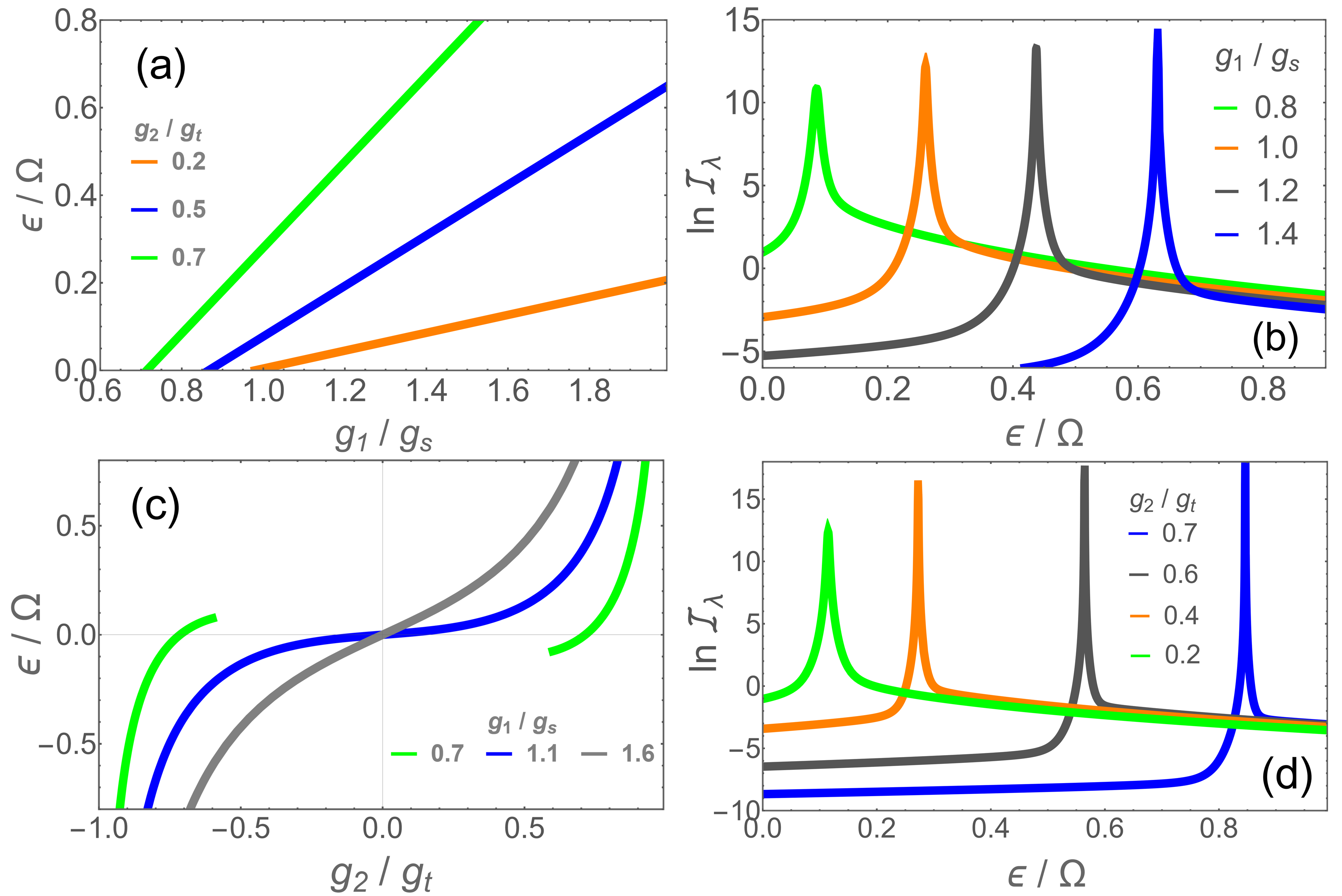}
\caption{(color online) (a) Phase boundaries as function of $g_{1}$
  at $g_{2}/g_{\mathrm{t}}=0.2$ (orange), $0.5$ (blue), $0.7$
  (green).
  (b) $\ln \I_\l$ for $\l=\epsilon/\Omega$ as function of $\epsilon $ for $g_{2}/g_{\mathrm{t}}=0.7$ and $(g_{1}/g_{\mathrm{s}},\omega /\Omega )=(0.8,0.1)$ (green), $(1.0,0.2)$ (orange), $(1.2,0.3)$ (dark gray) and $(1.4,0.4)$ (blue). (c) Phase boundaries as function of $g_{2}$ at $g_{1}/g_{\mathrm{s}}=0.7$ (green), $1.1$ (blue) and $1.6$ (gray). (d) $\I_\l$ as
function of $\epsilon$ for $g_{1}/g_{\mathrm{s}}=0.7$ at $(g_{2}/g_{\mathrm{t}},\omega /\Omega )=(0.2,0.2)$ (green), $(0.4,0.2)$
(orange), $(0.6,0.3)$ (dark gray) and $(0.7,0.4)$ (blue).
}
\label{fig-Fluxmeter}
\end{figure}

\section{Magnetometry}

The general Hamiltonian \eqref{ham} contains five parameters, all of which can be subjected to quantum metrology. We have focused, as an example, to the linear coupling $g_1$ but other parameters are interesting as well from a metrological point of view. The bias $\epsilon$ is of particular interest as it can be directly proportional to the intensity of external electric or magnetic fields in atomic and circuit-QED implementations, respectively. Using such a platform, it would be possible to construct a magnetometer analogous to a SQUID with enhanced precision. We have computed the QFI for $\l=\epsilon/\Omega$ in different parameter regions at finite frequencies. The results are shown in Fig.~\ref{fig-Fluxmeter}~(b) and (d) as function of the measured quantity $\epsilon$ for non-zero values of $g_2$ to take advantage of the non-linear coupling. Qualitatively we find the same features as for the previous case with $\l=g_1/\Omega$. In Fig.~\ref{fig-Fluxmeter}~(a) and (c) the phase boundaries are shown in the $\epsilon$/$g_1$ plane and the $\epsilon$/$g_2$ plane respectively. In each case, the whole range for $\epsilon$ can be attained by a phase boundary point if $g_1$ and $g_2$ are adjusted through a suitable adiabatic preparation process.

\section{Conclusions}

Via a study on the QFI and the gap of the nonlinear quantum Rabi model with
bias, we have compared the critical metrology provided by quantum phase transitions of different order. While the model with only linear coupling shows a transition of second order type with  a closing gap and smooth GS wave function, the transition of the model with additional non-linear coupling can be classified as first order featuring a finite gap and a discontinuous change of the GS wave function. The reason for this difference is the broken parity symmetry of the non-linear model which manifests itself in the GS wave function only at and above the critical point. In contrast, the linear, parity symmetric model has a GS changing smoothly across the transition. This leads to a dramatic increase of the QFI close to criticality in the non-linear case. Morover, the critical slowing down due to gap closing which extends the preparation time in the linear model is absent for non-linear coupling. A third advantage of the non-linear over the linear model is the possibility to avoid the slow-resonator limit as frequency rations of $\om/\Omega\sim0.1$ are sufficient to utilize the critical quantum enhancement of the measurement precision. This condition substantially eases the requirements for an experimental implementation. Finally, adding a standard bias term to the Hamiltonian extends the measurement range for the couplings to all realizable values because the critical point can be shifted by adjusting the bias.
On the other hand, one may construct a new type of magnetometer with critically enhanced precision if the bias itself is subjected to the measurement.

Therefore, the extension of the standard quantum Rabi model by including a non-linear coupling and the bias term may lead to a major improvement of quantum metrology in not just one but several respects.

\section*{Acknowledgements}

This work was supported by the National Natural Science Foundation of China
(Grant No. 11974151) and by the German Research Foundation (DFG) under Grant No. 439943572.

\bibliography{References}

\end{document}